\documentclass{rmaa}
\newcommand{\kms}{\mbox{$\,$km s$^{-1}$}}

\newcommand{\gr}{$^\circ$}

\newcommand{\nii}{[N~{\sc ii}]}

\newcommand{\oiii}{[O~{\sc iii}]}
\newcommand{\ha}{H$\alpha$}

\title{Enigmatic low-velocity jet-like features in planetary nebulae}

\author{Denise R. Gon\c calves
     \affil{Instituto de Astrof\'\i sica de Canarias} 
   Romano L. M. Corradi
     \affil{Isaac Newton Group of Telescopes} 
  \and Antonio Mampaso
     \affil{Instituto de Astrof\'\i sica de Canarias} 
}

\fulladdresses{
\item D.R. Gon\c calves and A. Mampaso: Instituto de Astrof\'\i sica de 
      Canarias, Calle Via L\'actea s/n, E-38200 La Laguna - Tenerife, Spain  
      (denise,amr@ll.iac.es)
\item R.M.L. Corradi: Isaac Newton Group of Telescopes, Apartado de Correos 
      321, E-38700, Sta. Cruz de la Palma, Spain (rcorradi@iac.es)
}

\shortauthor{Gon\c calves et al.}
\shorttitle{Low-velocity jet-like features in PNe}

\SetVolume{??} \SetFirstPage{1} \SetYear{2001}
\resumen{
Dentro de nuestro proyecto de investigaci\'on sobre las condiciones 
f\'\i sicas,
origen y evoluci\'on de las estructuras de baja ionizaci\'on en las Nebulosas
Planetarias, hemos identificado varios pares de estructuras tipo `jet'
altamente colimadas (Gon\c calves et al. 2001). Aunque son muy parecidas a los
`jets' reales, curiosamente presentan velocidades de expansi\'on muy 
peque\~nas, o
por lo menos no muy diferentes de las de las envolturas en las que se 
encuentran. Presentamos aqu\'\i \ nuestros datos sobre estos falsos `jets'
(Corradi et al. 1997, 1999) comp\'arandolos con los modelos te\'oricos 
existentes
sobre la formaci\'on de estructuras colimadas en NPs. Estos enigm\'aticos
sistemas `tipo jet' son dif\'\i ciles de encajar en los escenarios te\'oricos 
que tratan de las estructuras colimadas en NPs.
}

\abstract{
We are developing a project aimed at studying the physical properties, origin
and evolution of low-ionization structures in planetary nebulae. 
Within this project we have identified a number of
pairs of highly collimated low-ionization jet-like features (Gon\c calves et 
al. 2001). In spite of being very similar to real jets, they have the 
intriguing
property of possessing expansion velocities which are very low, or at least not
significantly different from, that of the shells in which they are embedded. In
this contribution we discuss our data on these fake jets (Corradi et al.
1997, 1999) and compare them with
 existing theoretical models for the formation of collimated structures in
PNe. These enigmatic jet-like systems are not easily accounted for within the 
theoretical scenarios that deal with collimated features in PNe.
}
      

\begin{document}

\maketitle

\section{Planetary nebulae with low-ionization structures}
The main components of a planetary nebula -- spherical and 
elliptical shells, bright rims, bipolar lobes, and halos -- are better
identified in the light of hydrogen and helium recombination lines, as
well as in the forbidden \oiii\ lines.  On the other hand, on usually smaller 
scales, the low-ionization 
structures (LISs) are features selected because they are prominent in
low-ionization lines (particularly \nii), fainter in \ha, and 
almost undetectable in \oiii. Of 50 planetary nebulae (PNe) known to contain  
LISs we find that $24\%$ have real jets and $18\%$ have jet-like LISs. 
The features we  call {\it jet-like} systems are highly collimated 
filaments:
 i) directed in the radial direction of the central star; ii) which appear in 
symmetrically opposite pairs; and, at variance with real {\it jets}, 
iii) move with 
relatively low expansion velocities (similar to those of the main shells 
in which they are embedded; see Table 1 of Gon\c calves et al. 2001 for 
references). Table 1 illustrates the main characteristics of 
jet-like LISs in contrast with those of the real jets. Those PNe that 
will be discussed in this contribution are distinguished in the table by 
bold letters.
\begin{table*}[tbp]
  \caption{Jets {\it versus} Jet-like LISs}
  \begin{center}
    \begin{tabular}{lccclccc}\hline\hline
     & Jets & & & & Jet-like Systems & & \\ \hline
     Object & Kinem. ages$^{\rm a}$ & Orientation$^{\rm b}$ & & Object 
     & Kinem. data & Orientation$^{\rm b}$ & Location$^{\rm c}$
     \\ \hline\hline 
      Fg 1 & older & 40\gr & & He 2-249 & yes & 0\gr & outside \\
      Hb 4 & younger & 90\gr & & {\bf IC 4593} & yes & - & outside \\
      He 2-186 & - & - & & {\bf K 1-2} & yes & -  & inside \\
      IC 4634 & - & - & & M 3-1$^{\rm d}$ & no & -  & outside \\
      K 4-47 & older & - & & NGC 6309$^{\rm d}$ & no & - & outside \\
      M 1-16 & older & - & & NGC 6751 & yes & - & outside \\
      NGC 3918 & coeval & - & & NGC 6881 & yes & 40\gr & outside \\
      NGC 6210 & younger & 90\gr & & NGC 7354$^{\rm d}$ & no & - & outside \\
      NGC 6543 & younger & 20\gr & & {\bf Wray 17-1} & yes & - & inside \\ 
      NGC 6884 & coeval & 50\gr & & & & \\
      NGC 6891 & coeval & 0\gr & & & & \\
      NGC 7009 & coeval & 30\gr & & & & \\
      \hline\hline
    \end{tabular}
   \end{center}
     {\footnotesize
     a) Quoted kinematic ages are those of the jets as compared to those 
     of the main nebular shells. \\
     b) Orientation column shows the approximate angle of the LISs with 
     respect to the major axis of the main shell. \\
     c) The apparent location of the tip of the jet-like system with respect 
     to the swept-up shell.\\
     d) M~3-1, NGC~6309, and NGC~7354 do not have kinematic data along the 
     LISs, thus they cannot be classified as jets. \\
     }
\end{table*}
\begin{figure*}[hbtp]
 \begin{center}
   %\includegraphics[width=8.4truecm]{fig1-above.jpeg}
   %\includegraphics[width=8.4truecm]{fig1-below.jpeg}
   %{\scriptsize
   \caption{ABOVE: \ha\ and \nii\ images of K~1-2. The left panel shows the 
diffuse elliptical \ha\ shell as well as the jet like-system. In the right 
panel is shown a closer view of the low-ionization features. The string of 
knots that forms the jet-like system and the filament (A, A') 
are indicated. Arrows identify the other LISs of this PN. BELOW: kinematics 
of K~1-2\ . Left panel: spectra along the jet-like system indicating  
two different velocity regimes (regions A and A'). Right panel: 
position--velocity plot of the measured velocities corrected for the
 systemic velocity. 
Filled and open circles represent \nii\ and \ha\ velocities, respectively. 
The dashed line shows the velocity field of the main shell.}
   \label{}
  \end{center}
\end{figure*}

\section{Our data for jet-like systems}

\subsection{K~1-2}
This PN is composed of a low surface brightness, elliptically shaped main 
nebula (see in Fig.1 the morphology of the \ha\ emission), in addition to a 
low-ionization jet-like system (at P.A.=--27\gr\ and well identified in both 
emission lines of Fig.1). K~1-2 also contains another system of knots 
that appears (projected on the plane of the sky) almost perpendicular to the 
first one.  

Its kinematics shows that the main elliptical shell has an expansion 
velocity of 25\kms\ (determined from the splitting of the \ha\ line emission), 
and that the inner regions of the P.A.=27\gr\ jet-like LIS share the expansion 
of the shell in which they are embedded. Note, however, that the tail 
of this system (the feature we called A' in Fig.1) has a peculiar velocity, 
which increases outwards -- accelerating from  25\kms\ at the shell 
position to 45\kms\ at its outermost region (see Corradi et al. 1999 for a 
complete discussion of this object). 

\subsection{IC~4593}
From the images in Fig.2 we see that IC~4593 has, in addition to its bright 
rim and asymmetrical shell, two systems of collimated LISs: a jet-like system 
(P.A.=--41\gr) and a pair of knots (P.A.=--118\gr). Resembling the case of 
K~1-2, also in this PN the two main pairs of LISs are almost mutually
perpendicular. The spectra of IC~4593 allow us to determine, from the velocity 
field in various lines, the heliocentric systemic velocity as being 
$21 \pm 3$~\kms\ (Fig.2). Along the jet-like system, 
labeled ABB' (P.A.=--41\gr) the measured 
velocities do not differ from that of the shell. Our analysis (Corradi et al. 
1997) suggests that 
the low radial velocities measured for the jet-like system are not due to 
projection effects -- as argued by other outhors (Harrington \& Borkowski 
2000) -- thus showing that this is a real low-velocity jet-like system.

\subsection{Wray~17-1}
From the detailed discussion of Corradi et al. (1999), Wray~17-1 
possesses a diffuse elliptical shell (\oiii\ emission) and two 
collimated low-ionization structures (\nii\ and ratio images of Fig.3). The 
jet-like system, at P.A.=--28\gr, has a direction (projected into the 
sky) completely different of that of the pair of knots (at P.A.=--103\gr). 
From the data on the kinematics of this object, also in Fig. 3, we determined 
the expansion velocity of the main elliptical shell as being 28~\kms . As is 
shown in the position--velocity diagram, the jet-like system (except for its
tail) is embedded in the 28~\kms\ shell and has radial velocities smaller or 
similar to those of the shell. Note, however, that the radial velocity of the 
tail A' is very peculiar; there is a velocity gradient of 
90~\kms\  between A' and A. 
Considering that A and A' are morphologically well aligned, we 
explain this gradient as an ionization effect (Corradi et al. 1999). 
\begin{figure*}[hbtp]
 \begin{center}
   \caption{ABOVE: \ha\ and \nii\ images of IC~4593. The left panel shows the 
inner rim and the asymmetrical shell, while the jet-like system is better 
identified on the right. The field of view is $45'' \times 45''$, with 
north at the top and east to the left. BELOW: \nii\ position--velocity plots 
superposed on the contour plots of the observed spectra along the jet-like 
system (ABB') and along the pair of knots (CC'). Circles represent the 
measured heliocentric velocities.
}
   \label{}
  \end{center}
\end{figure*}
\begin{figure*}[hbtp]
 \begin{center}
   \caption{ABOVE: \oiii , \nii, and  \oiii/\nii\ images of Wray~17-1. 
The main shell is easily seen in the left panel, while the jet-like system 
(AB, A'B') and the pair of knots (CD) are better seen in the middle
image. Note that the ratio image puts in evidence the high collimation of the 
LIS. BELOW: spectra and 
PV plot of Wray~17-1, along the jet-like features (P.A.=--28\gr). Open and 
filled circles represent \nii\ and \ha\ measured velocities, respectively, 
and are corrected for the systemic velocity of the nebula.}
   \label{}
  \end{center}
\end{figure*}

\section{Theoretical ideas for the origin of jet-like LIS's}
Theoretical models for the formation of collimated structures 
in planetary nebulae can be roughly grouped into two kinds: those based in 
the interacting stellar winds (ISW models) and those that consider the 
stellar and accretion-disk winds interaction (accretion-disk models). ISW  
models are related to the origin of collimated features in a single-star 
scenario, while the latter ones deal with processes occurring in binaries. The 
formation of jets by the ISW process was pointed out by Frank, Balick 
\& Livio (1996). Magnetohydrodynamical simulations of the ISW also proved to 
be successful in predicting some of the properties of the observed PNe 
with jets (Garc\'\i a-Segura 1997; Garc\'\i a-Segura  et al. 1999). In 
particular, the observed  
linear relationship between the jet expansion velocity and the distance from 
the center is one of the predictions of the latter models, while new results 
of the stagnation zone jets also predict this velocity behavior (see Steffen 
\& L\'opez, these proceedings). Jets originating from the interplay of the 
accretion disk and stellar winds in PNe have been investigated by, for 
instance, Soker \& Livio (1994), Livio \& Pringle (1997), and Reyes-Ru\'\i z 
\& L\'opez (1999). Generally,
what all the above models predict is the formation of i) two-sided highly 
collimated features with supersonic velocities larger than that of the main 
shell, ii) kinematic ages similar to (ISW models) or older than 
(accretion-disk models) that of the main shell, and iii) with an orientation 
that depends on the presence of precession, wobbling, or a misalignment 
between the main axis of the coupled system (Blackman, Frank \& Welch 2001; 
Garc\'\i a-Segura \& L\'opez 2001). 

On the other hand, the fake jets presented here are a bit more difficult 
to explain. Their main characteristic -- the one that makes them different 
from genuine jets -- is that their velocities are not peculiar with respect 
to the surrounding medium. Note that most jet-like systems are located 
(projected onto the sky) outside the swept-up shell of the host nebulae 
(as well as all the jets of Table 1. See also Table 2 of Gon\c calves et 
al. 2001). 
However, for two of the cases we show here the jet-like features 
are inside of the main shell. It seems that this property plays an 
important role when we try to compare observed properties and theoretical 
predictions. This is particularly true when we analyze certain details of the 
results found by R\'o$\dot{\rm z}$yczka \& Franco (1996) -- the first paper 
proposing magnetized interacting stellar winds to produce jets in PNe 
-- since their 
simulations show the formation of two pairs of well collimated flows (one 
pair per stagnation region, which are in the polar regions of the main 
shell), and in every pair one of the flows is of low velocity and directed 
toward the central star, and the other is of high velocity and expanding 
away from it. The stagnation zone jets of Steffen \& L\'opez (these 
proceedings) appear to produce low-velocity collimated features as well. 
Note, however, that in both models the low-velocity jet-like systems develop 
inside the 
swept-up shell, and that 
this is not the case for four of the six PNe containing 
jet-like systems that are well studied (see last column of Table 1). Another 
important 
issue is the lifetime of the low-velocity features that would arise 
during the evolution of the interacting winds of the above cited models. 
Unfortunately, this information
is not easily obtained from the observations since the useful 
comparison between kinematic ages of nebular components do not apply in 
this case. Therefore, the actual 
origin of the low-velocity jet-like system in PNe is clearly a challenge 
for further exploration.

\acknowledgments
It is with pleasure that DRG acknowledges A. Raga and L. Binette for the 
excellent 
organization of the meeting and for partial support. 
All of us wish to acknowledge the Spanish grant DGES PB97-1435-C02-01.
%\vspace*{4cm}

\end{document}